\documentclass[twocolumn,showpacs,amsmath,amssymb]{revtex4}
\usepackage{graphicx}
\usepackage{dcolumn}
\usepackage{bm}
\newcommand{\beq}{\begin{equation}}
\newcommand{\eeq}{\end{equation}}
\newcommand{\bea}{\begin{eqnarray}}
\newcommand{\eea}{\end{eqnarray}}
\newcommand{\eps}{\varepsilon}

\begin{document}

\author{N. V. Gnezdilov}
\affiliation{Kurchatov Institute, 123182 Moscow, Russia}

\author{E. E. Saperstein}
\affiliation{Kurchatov Institute, 123182 Moscow, Russia}

\author{S. V. Tolokonnikov}
\affiliation{Kurchatov Institute, 123182 Moscow, Russia}
\affiliation{Moscow Institute of Physics and Technology, 141700
Dolgoprudny, Russia.}

\title{Spectroscopic factors of magic and semimagic nuclei within the self-consistent theory of
finite Fermi systems.}


\begin{abstract}
A scheme is presented to find single-particle spectroscopic factors (SF) of magic and
semimagic nuclei within the self-consistent theory of finite Fermi systems (TFFS). In addition to the
energy dependence of the mass operator $\Sigma$ induced by the surface-phonon coupling  effects which
are commonly considered in this problem, the in-volume energy dependence of the
operator $\Sigma$ inherent in the self-consistent TFFS is also taken into account. This dependence
arises due to the effect of high-lying particle-hole excitations and persists in nuclear
matter. The self-consistent basis of the energy density functional method by Fayans {\it et al.} is
used. Both the surface and in-volume contributions to the SFs turned out to be of comparable
magnitude. The results for magic $^{40,48}$Ca and $^{208}$Pb nuclei and semimagic lead isotopes are
presented.

\end{abstract}

\maketitle

\section{Introduction}
A single-particle spectroscopic factor (SF) is an important characteristic of the single-particle
state \cite{BM1}. In the many-body approaches to nuclear theory \cite{Brown1,Giai} it is expressed in
terms of the so-called $Z$-factor, the residue of the single-particle Green function $G({\bf r_1},{\bf
r_2};\eps)$, \beq Z({\bf r}_1,{\bf r}_2)= \left(1- \left(\frac {\partial \Sigma({\bf r_1},{\bf
r_2};\eps)} {\partial \eps}\right)_{\eps= \eps_{\lambda}}\right)^{-1}, \label{Z-fac}\eeq where
$\eps_{\lambda}$ is the single-particle energy of the state under consideration and $\Sigma({\bf
r_1},{\bf r_2};\eps)$ is the mass operator, $G=G_0+G_0\Sigma G$. Here we use the notation of theory of
finite Fermi systems (TFFS) \cite{AB1}.

Usually the energy dependence of the mass operator is supposed to be the result of the coupling of
particles with the surface vibrations (``phonons'') \cite{Giai,Litv-Ring,Bort}. In
the self-consistent TFFS \cite{KhS}, the in-volume energy dependence of the mass
operator does also exist. It arises  due to the contribution to the operator
$\Sigma$ of high-lying collective excitations, the giant resonances, and non-collective particle-hole
excitations as well. As it is shown below, the contribution of this term to the $Z$-factor
(\ref{Z-fac}) is comparable to that from the  phonon coupling (PC) effects. The in-volume term depends
weakly on   the nucleus under consideration and the single-particle state $\lambda$, whereas the PC
one fluctuates strongly.

In this work, we deal with magic nuclei or with non-superfluid components of semi-magic nuclei. In
this case, simple formulas for PC corrections to  nuclear characteristics  in systems without pairing
\cite{EPL1,Levels,dmomYaF}  can be used. In practice,  we found proton and neutron SFs in magic
$^{40,48}$Ca and $^{208}$Pb nuclei and proton ones in semimagic lead isotopes. Calculations are
carried out with  the use of  the self-consistent basis generated  by  the energy density functional
(EDF) by Fayans {\it et al.} \cite{Fay} with the set of parameters DF3-a \cite{Tol-Sap}.

\section{Brief formalism}

The self-consistent TFFS approach \cite{KhS,AB2} was formulated for nuclei without pairing in terms of
the quasiparticle Lagrangian $L_q$, $L_q=\int d{\bf r}{\cal L}_q({\bf r})$, which is constructed to
generate the quasiparticle mass operator $\Sigma_q$. By definition  \cite{AB1,KhS}, the latter
coincides with the exact mass operator $\Sigma$ at the Fermi surface. In the mixed coordinate-momentum
representation, $\Sigma_q({\bf r},k^2;\eps)$ depends linearly on the momentum squared $k^2$ and on the
energy $\eps$ as well,\beq \Sigma_q({\bf r},k^2;\eps)= \Sigma_0({\bf r}) + \frac 1 {2m \eps^0_{\rm F}}
\hat{\bf k} \Sigma_1({\bf r})\hat{\bf k}+ \Sigma_2({\bf r}) \frac {\eps} {\eps^0_{\rm F}},
\label{Sigmaq}\eeq where $\eps^0_{\rm F}= k_{\rm F}^2/2m$ is the Fermi energy of nuclear matter. In
magic nuclei which are non-superfluid, the Lagrangian density ${\cal L}_q$ depends on three sorts of
densities $\nu_i({\bf r}),i=0,1,2$. The first two are analogs of the SHF densities $\rho({\bf
r}),\tau({\bf r})$, whereas the density $\nu_2({\bf r})$ is a new ingredient of the self-consistent
theory  being the single-particle energy density. In the explicit form, these densities are as
follows: \beq \nu_0({\bf r})= \sum n_{\lambda} \psi^*_{\lambda}({\bf r})\psi_{\lambda}({\bf r}),
\label{nu0}\eeq \beq \nu_1({\bf r})= \frac 1 {2m \eps^0_{\rm F}}\sum n_{\lambda}
\nabla\psi^*_{\lambda}({\bf r})\nabla\psi_{\lambda}({\bf r}), \label{nu1}\eeq
 \beq \nu_2({\bf r})= \frac 1 {\eps^0_{\rm F}}\sum
n_{\lambda} \eps_{\lambda} \psi^*_{\lambda}({\bf r})\psi_{\lambda}({\bf r}), \label{nu2}\eeq where
$\eps_{\lambda}$ and $n_{\lambda}$ are the quasiparticle  energies and occupation numbers. The
quasiparticle wave functions in Eqs. (\ref{nu0}) -- (\ref{nu2}) are normalized with the weight, \beq
\int d {\bf r}\,\psi^*_{\lambda}({\bf r})\,\psi_{\lambda'}({\bf r}) \,\left( 1- \Sigma_2({\bf
r})/\eps^0_{\rm F} \right) = \delta_{\lambda \lambda'} \label{norm}.\eeq
The usual density $\rho({\bf r})$ normalized to the total particle number is \beq \rho({\bf r}) =
\left( 1- \Sigma_2({\bf r})/\eps^0_{\rm F} \right) \nu_0 ({\bf r}) \label{ro}.\eeq
The $Z$-factor (\ref{Z-fac})  we are interested  is determined with the energy derivative \beq \frac
{\partial \Sigma_q({\bf r})} {\partial \eps}= \Sigma_2({\bf r})/\eps^0_{\rm F}. \label{dSig2dE} \eeq
In its turn, the  $\Sigma_2$ operator is \beq \Sigma_2({\bf r}) = \frac {\delta L_q} {\delta \nu_2}
\label{dL2dnu2}.\eeq  The simplest form of the corresponding term of the quasiparticle Lagrangian was
proposed in \cite{KhS,AB2} yielding  \beq \Sigma_2({\bf r})=C_0 \lambda_{02} \nu_0 ({\bf r})
\label{Sigma2}, \eeq where  $C_0=(dn/d\eps_{\rm F}^0)^{-1}= \pi^2/m p_{\rm F}^0 $ is the
usual TFFS normalization factor, the inverse density of states at the Fermi surface. The
dimensionless parameter $\lambda_{02}=-0.25$ is used  below, which is found in \cite{KhS} for the best
description of the single particle spectra of magic nuclei. It corresponds to the value $Z_0=0.8$ of
the $Z$-factor in  nuclear  matter. This value agrees with the one found in \cite{ST1,ST2} from the
dispersion  relation for the quantity $\partial \Sigma/\partial \eps$ in nuclear matter.

With the use of Eqs. (\ref{Z-fac}), (\ref{ro}), (\ref{dSig2dE}) and (\ref{Sigma2}), the $Z$-factor can
be expressed directly in terms of the usual density,  \beq Z_0({\bf r})=\frac 2 {1+
\sqrt{1-4C_0 \lambda_{02}\rho({\bf r})/\eps_{\rm F}^0}} \label{Z0}.\eeq
The self-consistent evaluation of the PC corrections to the mass operator in magic nuclei was studied
in \cite{Levels} with the analysis of single-particle energies. In this work we use the same approach
for spectroscopic factors.

\begin{figure}
\vspace{-5mm} \centerline {\includegraphics [width=80mm]{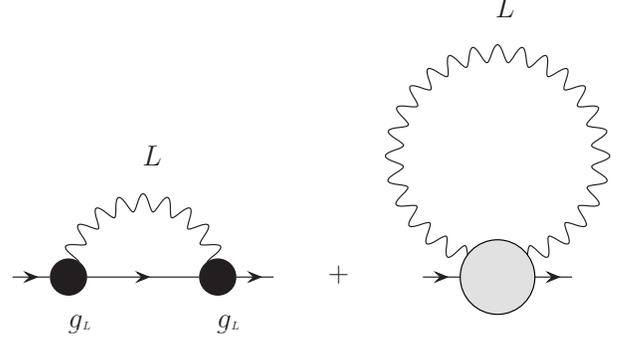}} \vspace{-2mm} \caption{PC corrections
to the mass operator. The gray blob denotes the ``tadpole'' term.} \label{fig:PC-cor}
\end{figure}

In accordance with Eq. (\ref{Z-fac}), the PC  contribution to the single-particle $Z$-factor is \beq
Z^{\rm PC} =
 \left( 1-\left.\frac{\partial\delta\Sigma^{\rm
PC}}{\partial\eps}\right|_ {\eps=\eps_{\lambda}}\right)^{-1} \label{ZPC} ,\eeq where $\delta
\Sigma^{\rm PC}$ is the PC correction to the quasiparticle mass operator $\Sigma_q$. Here we deal with
the perturbation theory in $\delta \Sigma$ operator with respect to the quasiparticle Hamiltonian
$H_0$.  As well as in \cite{EPL1,Levels,dmomYaF}, we use the so-called $g_L^2$-approximation which is,
as a rule,  applicable to magic nuclei. Here $g_L$ is the $L$-phonon creation amplitude. Actually,
$g_L^2$-correction to the mass operator contains two terms \cite{KhS}, \beq \delta\Sigma^{\rm PC} =
\delta\Sigma^{\rm pole}+\delta\Sigma^{\rm tad} \label{SigPC}, \eeq which are shown in Fig.
\ref{fig:PC-cor}.

In obvious symbolic notation, the pole diagram corresponds to
$\delta\Sigma^{\rm pole}=(g_L,DGg_L)$  where $D_L(\omega)$ is the
phonon $D$-function.  Explicitly one obtains \bea \delta\Sigma^{\rm
pole}_{\lambda\lambda}(\epsilon)&=&\sum_{\lambda_1\,M}
|\langle\lambda_1|g_{LM}|\lambda\rangle|^2 \nonumber\\
&\times&\left(\frac{n_{\lambda_1}}{\eps+\omega_L-
\eps_{\lambda_1}}+\frac{1-n_{\lambda_1}}{\eps-\omega_L -\eps_{\lambda_1}}\right), \label{dSig2} \eea
where $\omega_L$ is the excitation energy of the $L$-phonon and $n_{\lambda}=(0,1)$ stands for
occupation numbers.

The vertex $g_L$  in Eq. (\ref{dSig2}) reads \cite{AB1} \beq { g_L}(\omega)={{\cal F}} {A}(\omega) {
g_L}(\omega), \label{g_L} \eeq where $ A(\omega)=\int G \left(\eps + \omega/ 2 \right) G \left(\eps -
\omega/ 2 \right)d \eps/(2 \pi i)$ is the particle-hole propagator, $G(\eps)$ being the
 single-particle Green function.

As for the second, tadpole, term in Fig. \ref{fig:PC-cor}, it is   equal to  \beq \delta\Sigma^{\rm
tad}=\int \frac {d\omega} {2\pi i} \delta_L {g_L} D_L(\omega),\label{tad} \eeq where $\delta_L {g_L}$
can be found \cite{KhS} by variation of Eq. (\ref{g_L}) in the field of the $L$-phonon. We do not
focus on the term $\delta\Sigma^{\rm tad}$  because it does not depend on the energy
$\eps$. Hence, it does not contribute to $Z^{\rm PC}$. Vertices $g_L$ and the tadpole term are
considered in more detail in \cite{Levels}.

Thus, the energy derivative of the mass operator, which enters into (\ref{ZPC}),  is equal
to
 \bea \left. \frac{\partial\delta\Sigma^{\rm
pole}_{\lambda\lambda}(\eps)}{\partial\eps}\right|_{\eps=\eps_{\lambda}}
&=&-\sum_{\lambda'\,M}|<\lambda'|g_{LM}|\lambda>|^2 \nonumber\\
\times \left[\frac{n_{\lambda'}}{(\eps_{\lambda}+\omega_L- \eps_{\lambda'})^2}\right. &+&\left.
\frac{1-n_{\lambda'}} {(\eps_{\lambda} -\omega_L -\eps_{\lambda'})^2} \right] \label{ddSig}. \eea With
the use of Eqs. (\ref{Z-fac}) and (\ref{ZPC}), one obtains the total single-particle
$Z$-factor allowing for PC effects,  \beq Z_{\lambda} =  Z_{0{\lambda}} \left(1-{Z_0}_{\lambda}
\left.\frac{\partial}{\partial\eps}\delta\Sigma^{\rm
pole}_{\lambda\lambda}\right|_{\eps=\eps_{\lambda}}\right)^{-1} \label{Ztot} ,\eeq where
${Z_0}_{\lambda}=<\lambda|Z_0(\bf r)|\lambda>$.

The typical value of the denominators in the sum (\ref{ddSig}) is $\sim \eps_{\rm F}A^{-1/3}$.
However,  there are some cases when one of them is close to zero. This happens often in semimagic nuclei
where the excitation energy of the first $2^+$-state is $\omega_2\sim 1\;$MeV. In this case, the
single-particle states $\lambda,\lambda'$ which have the same parity and belong to the same shell can
possess close energies, so that  one gets  $|\eps_{\lambda} -\eps_{\lambda'}| \cong \omega_L$. As a
result, the simple perturbation theory used in (\ref{ddSig}) should be modified with separating the
couple of ``dangerous'' states  $(\lambda_0,\lambda'_0)$. Let us, e. g., have
$\eps_{\lambda_0} \cong \eps_{\lambda'_0} - \omega_L$. Then, for $\lambda=\lambda_0$, we extract the
term with $\lambda'=\lambda'_0$ in the sum (\ref{ddSig}). Let us denote the corresponding value of the
phonon $Z$-factor as $Z'$. The next step is the exact solution of the two-level problem of mixing the
states $|\lambda_0\rangle$ and $|\lambda'_0 + 2^+_1\rangle$. However, there are cases of almost exact
generation of these states with a trivial solution of this problem. Namely, both   mixing coefficients
are equal to $\sqrt{2}$. In this case, the phonon $Z$-factor is equal to \beq Z^{\rm
PC}_{\lambda_0}=\frac 1{\sqrt{2}} \, Z'_{\lambda_0}. \label{Zprim}\eeq

All the above formulas are written for magic nuclei in which the pairing does not exist. They are
 valid also for the normal component of a semimagic nucleus, with a sole exception. As far
as the second component is superfluid, the RPA-like Eq. (\ref{g_L}) should be replaced with the
QRPA-like TFFS equation for nuclei with pairing \cite{AB1}. In practice, we used the scheme developed
for superfluid nuclei in \cite{BE2}.

\begin{table}
\vspace{-5mm} \caption{Spectroscopic factors in $^{208}$Pb. Experimental data are taken from
\cite{bnl} and from \cite{Volya}, marked with $^a$.}
\begin{center}
\begin{tabular}{cccccc}
\noalign{\smallskip}\hline\noalign{\smallskip}

$\lambda$ & $\eps_{\lambda}$ &  $Z_0$ & $Z^{\rm PC}$ & $Z_{\rm tot}$ & $S_{\rm exp}$\\

\noalign{\smallskip}\hline\noalign{\smallskip} \noalign{\smallskip}

$3d_{3/2}^n$ &  -0.709 & 0.852  &  0.879  &  0.763 & 0.74 \\
$4s_{1/2}^n$ &  -1.080 & 0.848  &  0.895  &  0.771 & 0.8 \\
$2g_{7/2}^n$ &  -1.091 & 0.849  &  0.886  &  0.765 & 0.9 \\
$3d_{5/2}^n$ &  -1.599 & 0.848  &  0.873  &  0.755 & 0.85 \\
$1j_{15/2}^n$&  -2.167 & 0.830  &  0.618  &  0.549 & 0.53 \\
$1i_{11/2}^n$&  -2.511 & 0.838  &  0.945  &  0.799 & 0.9 \\
$2g_{9/2}^n$ &  -3.674 & 0.833  &  0.882  &  0.749 & 0.8 \\
$3p_{1/2}^{-n}$ &  -7.506 & 0.834  &  0.926  &  0.782 & 1.1,  0.9$^a$ \\
$3p_{3/2}^{-n}$ &  -8.363 & 0.822  &  0.913  &  0.762 & 0.96, 0.88$^a$ \\
$2f_{5/2}^{-n}$ &  -8.430 & 0.828  &  0.923  &  0.775 & 0.88, 0.6$^a$ \\
$1i_{13/2}^{-n}$&  -9.411 & 0.822  &  0.902  &  0.755 & 0.49, 0.91$^a$ \\
$2f_{7/2}^{-n}$ & -10.708 & 0.816  &  0.567  &  0.503 & 0.55, 0.95$^a$ \\
$1h_{9/2}^{-n}$ & -11.009 & 0.820  &  0.892  &  0.746 & 0.56, 0.98$^a$  \\
\noalign{\smallskip}\hline\noalign{\smallskip} \noalign{\smallskip}

$3p_{3/2}^p$ &  -0.249 & 0.853  &  0.690  &  0.617 & 1.03 \\
$2f_{5/2}^p$ &  -0.964 & 0.850  &  0.812  &  0.710 & 1.15 \\
$1i_{13/2}^p$&  -2.082 & 0.836  &  0.741  &  0.646 & 0.88 \\
$2f_{7/2}^p$ &  -3.007 & 0.842  &  0.859  &  0.740 & 1.18 \\
$1h_{9/2}^p$ &  -4.232 & 0.837  &  0.958  &  0.807 & 0.95 \\
$3s_{1/2}^{-p}$ &  -7.611 & 0.838  &  0.929  &  0.787 & 0.85$^a$ \\
$2d_{3/2}^{-p}$ &  -8.283 & 0.827  &  0.937  &  0.783 & 0.9$^a$ \\
$1h_{11/2}^{-p}$&  -8.810 & 0.832  &  0.931  &  0.784 & 0.88$^a$ \\
$2d_{5/2}^{-p}$ &  -9.782 & 0.826  &  0.711  &  0.618 & 0.65$^a$ \\
$1g_{7/2}^{-p}$ & -11.735 & 0.820  &  0.423  &  0.387 & 0.52$^a$ \\
\noalign{\smallskip}\hline\noalign{\smallskip}
\end{tabular}
\end{center}
\label{tab:Pb208}
\end{table}

\begin{table}
\caption{ Spectroscopic factors in $^{40}$Ca. Experimental data are taken from \cite{bnl}.}

\begin{tabular}{cccccc}
\noalign{\smallskip}\hline\noalign{\smallskip}

$\lambda$ & $\eps_{\lambda}$ &  $Z_0$ & $Z^{\rm PC}$ & $Z_{\rm tot}$ & $S_{\rm exp}$\\

\noalign{\smallskip}\hline\noalign{\smallskip} \noalign{\smallskip}

$1f_{5/2}^n$ &  -2.124 & 0.910  &  0.947  &  0.866 & 0.9 \\
$2p_{1/2}^n$ &  -3.729 & 0.917  &  0.934  &  0.861 & 0.7 \\
$2p_{3/2}^n$ &  -5.609 & 0.905  &  0.916  &  0.836 & 0.91 \\
$1f_{7/2}^n$ &  -9.593 & 0.893  &  0.947  &  0.850 & 0.77 \\
$1d_{3/2}^{-n}$ & -14.257 & 0.871  &  0.965  &  0.844 & 0.94 \\
$2s_{1/2}^{-n}$ & -15.780 & 0.875  &  0.930  &  0.821 & 0.82 \\
$1d_{5/2}^{-n}$ & -19.985 & 0.868  &  0.969  &  0.844 & 0.9 \\
\noalign{\smallskip}\hline\noalign{\smallskip}
$1f_{5/2}^p$ &   4.359 & 0.921 &   0.963  &  0.890 & 0.33 \\
$2p_{1/2}^p$ &   2.456 & 0.935 &   0.950  &  0.891 & 0.75 \\
$2p_{3/2}^p$ &   0.936 & 0.916 &   0.966  &  0.887 & 0.94 \\
$1f_{7/2}^p$ &  -2.678 & 0.896 &   0.960  &  0.864 & 1.12 \\
$1d_{3/2}^{-p}$ &  -7.264 & 0.874 &   0.966  &  0.848 & 0.93 \\
$2s_{1/2}^{-p}$ &  -8.663 & 0.878 &   0.931  &  0.825 & 0.87 \\
$1d_{5/2}^{-p}$ & -12.856 & 0.870 &   0.969  &  0.846 & 0.83 \\

\noalign{\smallskip}\hline\noalign{\smallskip}
\end{tabular}
\label{tab:Ca40}
\end{table}

\section{Calculation results}

Before we start to discuss the calculation results for SFs, a comment on the
``experimental'' data should be made. We took the word ``experimental'' in quotes as this term could be
used for the corresponding data, say, in the database \cite{bnl}, rather conventionally. Indeed, ``a
lot of theory'' is used to extract SFs from the experimental cross sections. Typically, SFs are found
from the analysis of $(d,p)$ or $(d,n)$ reactions within the Distorted Wave Born Approximation (DWBA).
Such a procedure was convincingly criticized in Ref. \cite{DWBA1} titled ''Non-observability of
Spectroscopic Factors''. Arguments are given that the ansatz for the energy dependence of the optical
potential used in the DWBA calculation  affects the SF values more than the measured
cross-sections. In such a situation, it is concluded in \cite{DWBA1} that it is more reasonable to put
attention to a tendency in variance of SF values found in a series of experiments, e. g.  for a chain
of isotopes, than for their absolute values. The reaction $(e,e'p)$ could be also used for extracting
SFs, being simpler for the analysis than deuteron induced reactions.  However, as the analysis in Ref.
\cite{DWBA2} shows, even in this most simple reaction there are doubts in a possibility to extract the
occupation numbers consistently. Below, we will omit the quotation marks in the word experimental
keeping them in mind.

 It should be stressed that our calculations do  not contain any new parameters. The set
DF3-a parameters \cite{Tol-Sap} of the Fayans functional proved to be rather successful in describing
various nuclear characteristics including nuclear charge radii \cite{Sap-Tol}, the excitation energies
and $BE2$ values for $2_1^+$ states in lead and tin isotopes \cite{BE2}, the quadrupole moments of odd
nuclei \cite{QEPJ}, PC corrections to magnetic moments \cite{EPL1,dmomYaF}, and single-particle levels
of magic nuclei \cite{Levels}. As to the parameter $\lambda_{02}$ governing the value of $Z_0$ in
nuclear matter, it was found in \cite{KhS} from the analysis of single-particle spectra in magic
nuclei. The value was confirmed \cite{ST1,ST2} with the analysis of the dispersion relation for the
quantity $\partial \Sigma/\partial \eps$ in nuclear matter. In its turn, the latter was resolved in
terms of the known Landau--Migdal parameters.

Let us begin  the analysis  from the double magic $^{208}$Pb, $^{40}$Ca and $^{48}$Ca nuclei, tables
1--3, with the greatest amount of the  experimental  data on SFs. We took the data mainly from the
database \cite{bnl}. The latter typically contains several values for the state and the nucleus under
consideration taken from different original works which sometimes are significantly different, in
agreement with \cite{DWBA1}. As a rule, we chose the freshest one, just to have some formal how-to-do recipe.  
In several cases we used also the experimental SFs from \cite{Volya} where some original
experimental works were used in addition to \cite{bnl}. Each table contains three values for the
$Z$-factor which are the theoretical predictions for the SF. ${Z_0}$ takes into account only in-volume
energy dependence of the quasiparticle mass operator (\ref{Sigmaq}) within the self-consistent TFFS. $
Z^{\rm PC}$ is the SF originating only from the PC contributions. At last, $ Z_{\rm tot}$ is the total
SF which takes into account both effects via Eq. (\ref{Ztot}).

\begin{table}
\caption{Spectroscopic factors in $^{48}$Ca. Experimental data are taken from \cite{bnl}.}
\begin{center}
\begin{tabular}{cccccc}
\noalign{\smallskip}\hline\noalign{\smallskip}

$\lambda$ & $\eps_{\lambda}$ &  $Z_0$ & $Z^{\rm PC}$ & $Z_{\rm tot}$ & $S_{\rm exp}$\\

\noalign{\smallskip}\hline\noalign{\smallskip} \noalign{\smallskip}

$1g_{9/2}^n$ &  0.836 & 0.915  &  0.796  &  0.741 & 0.47 \\
$2p_{1/2}^n$ & -3.890 & 0.880  &  0.773  &  0.699 & 1.03 \\
$2p_{3/2}^n$ & -5.784 & 0.854  &  0.939  &  0.809 & 0.97 \\
$1f_{7/2}^{-n}$ & -9.488 & 0.870  &  0.965  &  0.844 & 0.85 \\
\noalign{\smallskip}\hline\noalign{\smallskip} \noalign{\smallskip}

$2p_{1/2}^p$ & -3.549 & 0.883  &  0.648  &  0.596 & 0.88 \\
$1f_{5/2}^p$ & -4.048 & 0.890  &  0.873  &  0.788 & - \\
$2p_{3/2}^p$ & -4.731 & 0.880  &  0.604  &  0.558 & 0.54 \\
$1f_{7/2}^p$ & -9.909 & 0.865  &  0.899  &  0.789 & 0.91 \\
$2s_{1/2}^{-p}$ &-15.098 & 0.853  &  0.915  &  0.791 & 0.92 \\
$1d_{3/2}^{-p}$ &-16.172 & 0.858  &  0.917  &  0.796 & 1.22 \\

\noalign{\smallskip}\hline\noalign{\smallskip}
\end{tabular}
\end{center}
\label{tab:Ca48}
\end{table}

\begin{table}
\caption{Spectroscopic factors of proton states in semimagic isotopes $^{204,206}$Pb. Experimental
data are taken from \cite{bnl}.
* marks the case when the $Z^{\rm PC}$ was found from Eq. (\ref{Zprim}).}

\begin{tabular}{ccccccc}
\noalign{\smallskip}\hline\noalign{\smallskip}

$A$ & $\lambda$ & $\eps_{\lambda}$ &  $Z_0$ & $Z^{\rm PC}$ & $Z_{\rm tot}$ & $S_{\rm exp}$\\

\noalign{\smallskip}\hline\noalign{\smallskip} \noalign{\smallskip}

    & $1i_{13/2}^{p}$  &  -1.666 & 0.836 &   0.821  &  0.707 & 0.90 \\
    & $2f_{7/2}^{p}$   &  -2.513 & 0.842 &   0.817  &  0.708 & 0.82 \\
206 & $1h_{9/2}^{p}$   &  -3.818 & 0.838 &   0.880  &  0.752 & 0.90 \\

    & $3s_{3/2}^{-p}$  &  -7.178 & 0.839 &   0.926  &  0.786 & 0.64 \\
    & $2d_{3/2}^{-p}$  &  -8.283 & 0.827 &   0.327  &  0.306 & 0.49 \\
    & $1h_{11/2}^{-p}$ &  -8.416 & 0.832 &   0.859  &  0.732 & 0.61 \\

\noalign{\smallskip}\hline\noalign{\smallskip}

    & $1i_{13/2}^{p}$  &  -1.210 & 0.836 &   0.683  &  0.602 & 0.82 \\
    & $2f_{7/2}^{p}$   &  -2.010 & 0.843 &   0.702  &  0.620 & 0.53 \\
204 & $1h_{9/2}^{p}$   &  -3.356 & 0.838 &   0.773  &  0.673 & 0.90 \\

    & $3s_{3/2}^{-p}$  &  -6.717 & 0.839 &   0.866  &  0.743 & 0.7 \\
    & ${2d_{3/2}^{-p}}^*$  &  -7.505 & 0.827 &   0.575  &  0.513 & 0.46 \\
    & $1h_{11/2}^{-p}$ &  -7.977 & 0.833 &   0.739  &  0.644 & 0.68 \\

\noalign{\smallskip}\hline\noalign{\smallskip}

\end{tabular}
\label{tab:Pb}
\end{table}

 As it was discussed above,  the experimental values of SFs   have large uncertainties. This is
particularly evident in several cases with $S_{\rm exp}\geq 1$.  This permits to suspect that in many
other cases the normalization of SFs could also be doubtful.  Tables 1--3 contain  four  such ``bad''
cases for $^{208}$Pb, one for $^{40}$Ca, and two for $^{48}$Ca. In addition to direct experimental
inaccuracies, these uncertainties appear because of the way to extract SFs from the data on the
one-nucleon transfer reactions  using the DWBA. Dropping at a time the principle objections
\cite{DWBA1} against the method itself, there is a lot of practical uncertainties in the procedure. In
particular,  the parameters of optical potentials used in the DWBA calculations are not known
sufficiently well, especially the spin-orbit ones.

\begin{figure}
\vspace{-2mm} \centerline {\includegraphics [width=80mm]{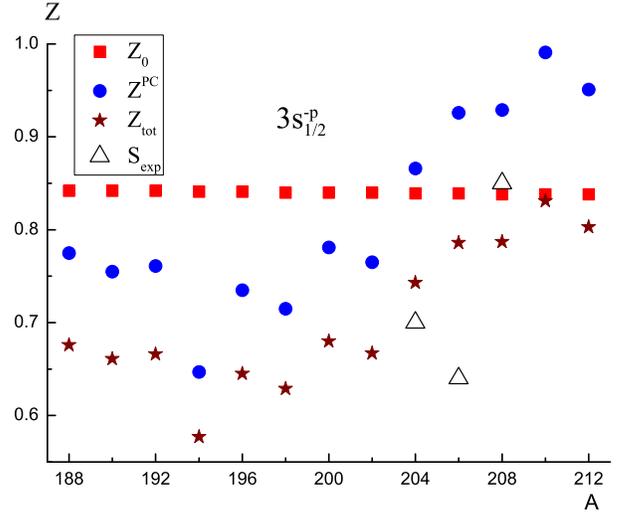}} \vspace{-2mm} \caption{(Color online)
Spectroscopic factors of the proton hole $3s_{1/2}^{-p}$ state in Pb isotopes.} \label{fig:s1/2}
\end{figure}

\begin{figure}
\vspace{-2mm} \centerline {\includegraphics [width=80mm]{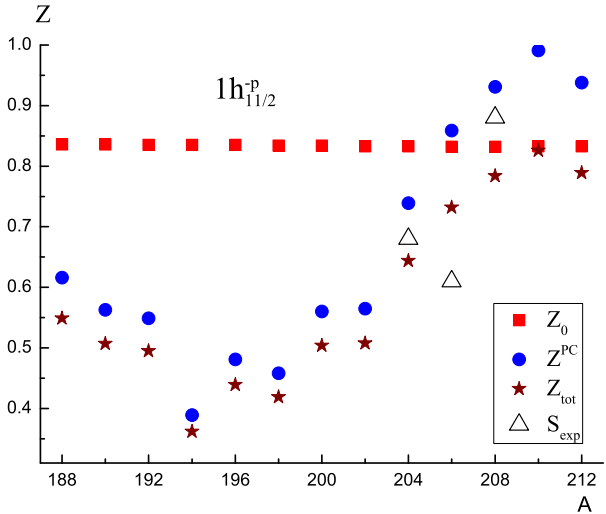}} \vspace{-2mm} \caption{(Color online)
Spectroscopic factors of the proton hole $1h_{11/2}^{-p}$ state in Pb isotopes.} \label{fig:h11/2}
\end{figure}

To understand which of three theoretical values for the $Z$-factor should be compared to the
experimental SF, let us  for a moment forget about principle problems with the DWBA approach and
imagine that there exists a perfect theory for describing reactions $(d,p)$ and $(d,n)$.  At small
energy of the deuteron projectile and, correspondingly, small momentum transfer $q\ll k_{\rm F}$, the
wave functions of the outgoing nucleon and the one absorbed with the target nucleus ``have time'' to
adapt to the mean field of the nucleus which, according to TFFS \cite{AB1}, involves the factor $Z_0$:
$U({\bf r})=Z_0({\bf r})\Sigma_q({\bf r})$. Therefore the SF extracted from such a reaction includes
only the spread  of the single-particle state under consideration over near-lying many-particle
configurations, those with PC contributions being the most important. In this case, the value of
$Z_{\lambda}^{\rm PC}$ should be compared to $S_{\rm exp}$. Another situation takes place in the case
of large values of the deuteron energy and momentum transfer $q \gg k_{\rm F}$. In this limit, the
field created by the projectile feels ``bare'' nucleons. The weight of a bare nucleon in
the state under consideration is equal to ${Z_0}_{\lambda}$. As a result, the experimental SF should
be compared with the total $Z$-factor $Z^{\rm tot}_{\lambda}$.

For relevant choice between    $ Z^{\rm PC}$ and $ Z_{\rm tot}$ values, it is necessary to make a detailed analysis of the experimental details and the theoretical method used for description of the reaction
data. It is  practically impossible  as far as the most of the experiments presented  in \cite{bnl}
are rather updated. In absence of such analysis, we are forced to use  some intermediate value between
the phonon and total SFs to compare with $S_{\rm exp}$. It should be mentioned also that the data on
SF in \cite{bnl} do  not contain as a rule the experimental bars. This stresses that the authors of the
original works   consider the accuracy of extracting  SF as rather poor.
 In such a situation, it is hardly reasonable to make a  detailed comparison with the data. If we exclude
 from the analysis four cases with $S_{\rm exp} \geq 1$, a rough analysis shows that the experimental value
 is, as a rule, in the interval between $ Z_{\rm tot}$ and $ Z^{\rm PC}$. The cases of preference
 of each limit value are divided approximately in half.   As in \cite{EPL1,Levels,dmomYaF},
 the PC $Z$-factor in $^{208}$Pb   was found by taking into account nine phonons: $3^-$,
two phonons $5^-$ and six phonons of positive parity,  $2^+_{1,2}$, $4^+_{1,2}$, and $6^+_{1,2}$. Just
as in calculations of the PC contributions to the single-particle spectra  \cite{Levels}, the
contribution of the $3^-$ phonon as a rule dominates. However, sometimes the total contribution of
other phonons is comparable to the one of $3^-$. Let us recall that $ Z^{\rm PC}$ is found in the
$g_L^2$ approximation which is valid provided we get $1-Z^{\rm PC} \ll 1$. This approximation
evidently fails for the state $1g_{7/2}^{-p}$, see the last line in Table 1. This explains why the
corresponding values $ Z^{\rm PC}$ and $ Z_{\rm tot}$ are too small to explain the datum. Next
correction in $g_L^2$ should make $ Z^{\rm PC}$ greater, closer to $S_{\rm exp}$. Another situation
takes place for the state $1i_{13/2}^p$. In this case, the $g_L^2$ approximation is valid and we
believe in the theoretical result. We suspect that, just as for its neighbors in the table, it is
worth to recall the quotation marks in the word ``experimental''.

In the $^{40}$Ca nucleus, Table 2, the value of  $ Z_{\rm tot}$ agrees as a rule better with the data.
There is only one third of cases where the $ Z^{\rm PC}$ value happens to be better. In this nucleus the
single $3^-$ phonon ($\omega_{3-}=3.335\;$MeV, $\omega_{3-}^{\rm exp}=3.737\;$MeV)  contributes to PC
effects. Although it is strongly collective [$BE3=22.9\;$WU, $BE3^{\rm exp}=27(4)\;$WU], $ Z^{\rm PC}$
values turn out to be close to unity, as all energy denominators in (\ref{ddSig}) are very large because of
the large values of both proton and neutron magic gaps. In the $^{48}$Ca nucleus, just as $^{208}$Pb, the
cases of the preferred $ Z^{\rm PC}$ and $ Z_{\rm tot}$ values are approximately bisected. In this
nucleus, two phonons, $3^-$ ($\omega_{3-}=4.924\;$MeV, $\omega_{3-}^{\rm exp}=4.507\;$MeV)  and $2^+$
 ($\omega_{2+}=E=3.576\;$MeV, $\omega_{2+}^{\rm exp}=3.832\;$MeV), contribute  to $ Z^{\rm
PC}$ approximately equally.  In this nucleus, the difference $(1-Z^{\rm PC})$ is, as a rule, bigger
than in $^{40}$Ca as now the neutron energy denominators of Eq. (\ref{ddSig}) are much smaller. In Ca
isotopes, there are two cases, $1f_{5/2}^p$-state in $^{40}$Ca and $1g_{9/2}^n$-state in $^{48}$Ca,
where a strong disagreement with the data is observed. In both cases, just as for the
$1i_{13/2}^p$-state in $^{208}$Pb, we believe that our predictions are more reliable.

In semimagic nuclei, the situation is essentially different due to the presence of the low-lying $2^+$
phonon which is strongly collective and dominates in all PC effects. As a result, the difference of
$Z^{\rm PC}$ from unity is, as a rule, bigger than in magic nuclei. It makes sometimes questionable
application of the perturbation theory used above for describing the PC effects. We consider the
proton SFs in the chain of the even semimagic lead isotopes $^{188-212}$Pb. Unfortunately, the data on
SFs in these nuclei are scarce. Therefore Table 4 contains two nuclei only, $^{204}$Pb and $^{206}$Pb,
for which the database \cite{bnl} contains some data on SFs. Predictions for all the chain are
displayed in figs. 2 and 3 for the states $3s_{1/2}^{-p}$ and $1h_{11/2}^{-p}$, respectively.. They
are chosen because, firstly, their experimental SFs are known in the nuclei listed above and, secondly,
there is a hope for measurement of their SFs  in other nuclei.  Indeed, the state $3s_{1/2}^{-p}$
there is right at the Fermi surface and therefore is readily populated in one-nucleon transfer
reactions. As to the  $1h_{11/2}^{-p}$ state, it has a high value of the angular momentum which makes
the cross section of its excitation larger. It is worth to note that there are three cases on fig. 3
with $ Z^{\rm PC} \preceq 0.5$, where the $g_L^2$-approximation is not valid. They are included just
for completeness of the illustration.

\section{Conclusion}
The self-consistent TFFS \cite{KhS} is used to calculate the spectroscopic factors of double magic
nuclei $^{40}$Ca, $^{48}$Ca and $^{208}$Pb  and the chain of semimagic lead isotopes $^{188-212}$Pb as
well. In this approach, the SF is related to the $Z$-factor, the residue of the single-particle Green
function in the single-particle pole $ \lambda$. The self-consistent basis is used of the Fayans EDF
\cite{Fay} with the set of parameters DF3-a \cite{Tol-Sap}. The total $Z$-factor (\ref{Ztot}) contains
the PC term induced by the low-lying surface excitations, similarly to that considered in Refs.
\cite{Giai,Litv-Ring,Bort}. In addition, it involves the in-volume component due to the energy
dependence of the quasiparticle mass operator inherent to the self-consistent TFFS. It is induced by
the high-lying in-volume virtual nuclear excitations, the spin-isospin ones playing the main role
\cite{ST1,ST2}. The in-volume and PC contributions to the total $Z$-factor are, as a rule, comparable
to each other. The in-volume effect depends weakly on the nucleus under consideration and on the state
$ \lambda$. On the contrary, the PC contribution fluctuates strongly depending on the nucleus and the
state $ \lambda$ we deal with.

The experimental SF should be related to the phonon $Z$-factor $Z^{\rm PC}$ in the case of low-energy
reactions with the momentum transfers $q\ll k_{\rm F}$. In the opposite case of $q\gg k_{\rm F}$, the
total $Z$-factor $Z_{\rm tot}$ should be used. In an intermediate situation, the experimental SF
should be compared to some intermediate value between $Z^{\rm PC}$  and $Z_{\rm tot}$. Keeping in mind
large uncertainties in the absolute values of the ``experimental'' SFs \cite{DWBA1,DWBA2}, a direct
comparison for each individual case is not much informative. In such a situation, we limit ourselves
to the comparison of a tendency in our calculations with the database \cite{bnl}, finding a satisfactory overall
agreement.

\section{Acknowledgment}

 The work was partly supported  by the Grant NSh-932.2014.2 of the Russian Ministry
for Science and Education, and by the RFBR Grants 12-02-00955-a, 13-02-00085-a, 13-02-12106-ofi\_m,
14-02-00107-a, 14-02-31353 mol\_a.

{}

\end{document}